\documentclass[fleqn,10pt]{wlscirep}
\usepackage[utf8]{inputenc}
\usepackage[T1]{fontenc}

\usepackage{bm}
\usepackage{graphicx}
\usepackage{dsfont}

\usepackage{mathrsfs}

\usepackage{framed}
\usepackage{multirow}
\usepackage[all]{xy}

\usepackage{framed}
\usepackage{comment}
\usepackage{here}                    
\usepackage{latexsym}		     
\usepackage{amsmath}     
\usepackage{amsfonts}  
\usepackage{amssymb}
\usepackage{amsthm}
\usepackage{dcolumn}
\usepackage{color}
\usepackage{mathrsfs}
\usepackage{dsfont}

\renewcommand{\phi}{\varphi}

\newcommand{\be}{\begin{equation}}
\newcommand{\ee}{\end{equation}}
\newcommand{\bea}{\begin{eqnarray}}
\newcommand{\eea}{\end{eqnarray}}

\renewcommand{\phi}{\varphi}

\title{Quantum key distribution with post-processing driven by physical unclonable functions}

\author[1,2,*]{Georgios M. Nikolopoulos}
\author[3]{Marc Fischlin}
\affil[1]{Institute of Electronic Structure and Laser, Foundation for Research and Technology-Hellas (FORTH), GR-70013 Heraklion, Greece}
\affil[2]{Institut f\"ur Angewandte Physik, Technische Universit\"at Darmstadt, D-64289 Darmstadt, Germany}
\affil[3]{Cryptoplexity, Technische Universit\"at Darmstadt, D-64289 Darmstadt, Germany}

\begin{abstract}
Quantum key-distribution protocols allow two honest distant parties to establish a common truly random secret key in the presence of powerful adversaries, provided that the two users share beforehand a short secret key. This pre-shared secret key is used mainly for authentication 
purposes in the post-processing of classical data that have been obtained during the quantum communication stage, and it prevents a man-in-the-middle attack. 
The necessity of a pre-shared key is usually considered as the main drawback of quantum 
key-distribution protocols, which becomes even stronger for large  networks involving more that two users. Here we discuss the conditions under which physical unclonable function can be integrated 
in currently available quantum key-distribution systems, in order to 
facilitate the generation and the distribution of the necessary pre-shared key, with the smallest possible cost in the security of the systems. Moreover, the integration of physical unclonable functions in quantum key-distribution networks  allows for real-time 
authentication of the devices that are connected to the network.  
\end{abstract}
\begin{document}

\flushbottom
\maketitle

\thispagestyle{empty}

\section*{Introduction}

Quantum key distribution (QKD) is the most mature quantum technology 
\cite{qkd_review1,qkd_review2,qkd_review3,qkd_review4,qkd_review5,qkd_review6,qkd_review7,qkd_review8,qkd_book}. 
Different QKD systems are available on the market by various companies, 
whereas national and international initiatives aim at testing the integration of the available QKD 
systems into existing communication infrastructures, as well  their operation 
in various use cases under realistic conditions. 

\begin{figure}
\centering\includegraphics[width=8.5cm]{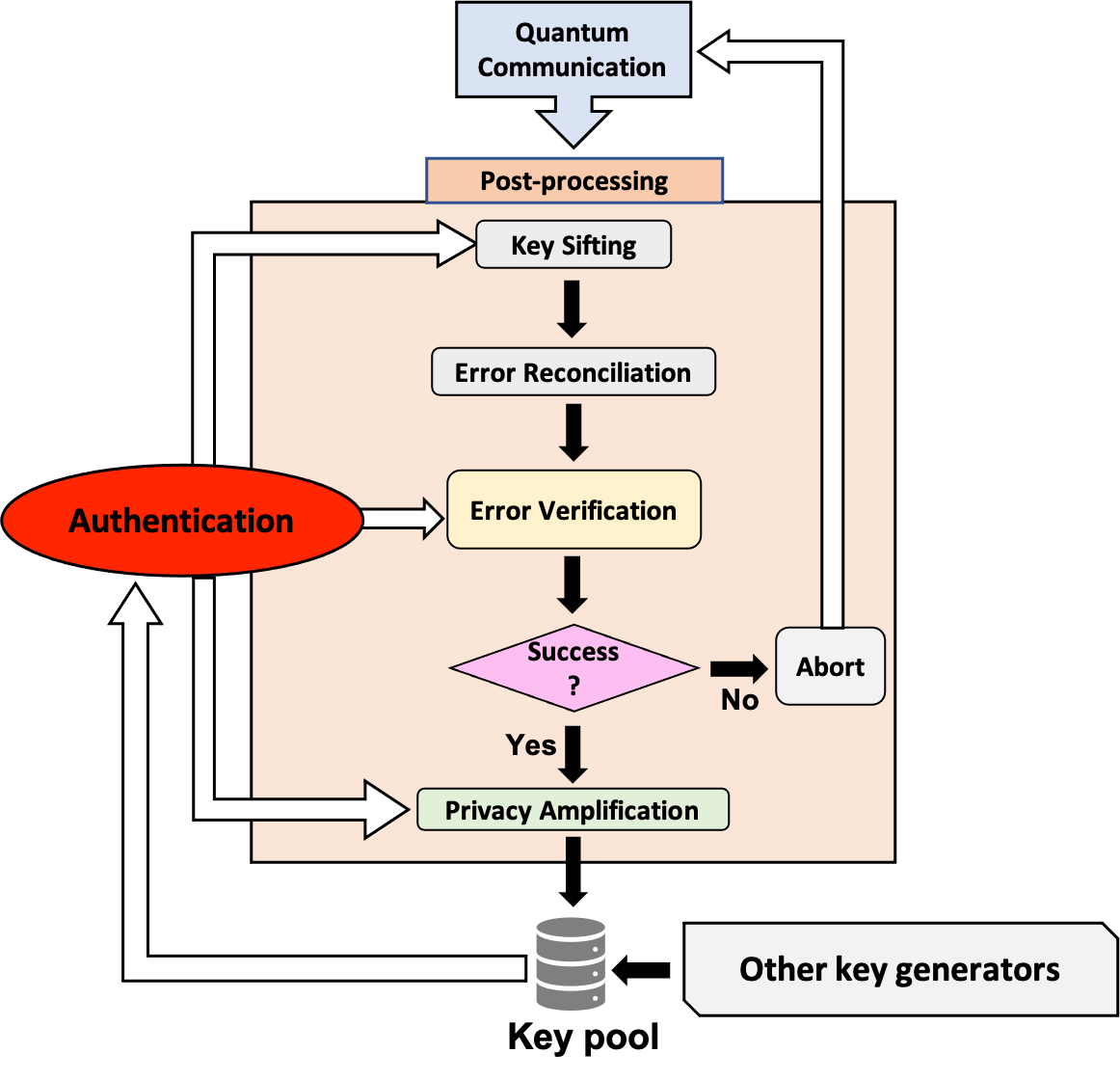}
\caption{Schematic presentation of the main steps and the flow of data 
in a QKD protocol. }
\label{figure1}
\end{figure}

Typically, the establishment of a secret key between two honest users (Alice and Bob) pertains to a quantum-communication and a post-processing 
stage (see Figure \ref{figure1}). In the first stage, Alice encodes random bits on non-orthogonal photonic states, which are sent to Bob over a quantum channel. Bob measures each received signal and records the outcomes. Subsequently, the 
two users post-process publicly their classical data, in order to obtain a final information-theoretically secure secret key. The main steps involved in the post-processing stage are key sifting, error reconciliation, error verification and privacy amplification. 
A successful implementation of all of these steps, without any detected failure, is expected to lead to a final secret key. However, for any of these steps, there is always a probability for undetected failure, in which case the security of the protocol is compromised. 

\begin{figure}
\centering\includegraphics[width=8.5cm]{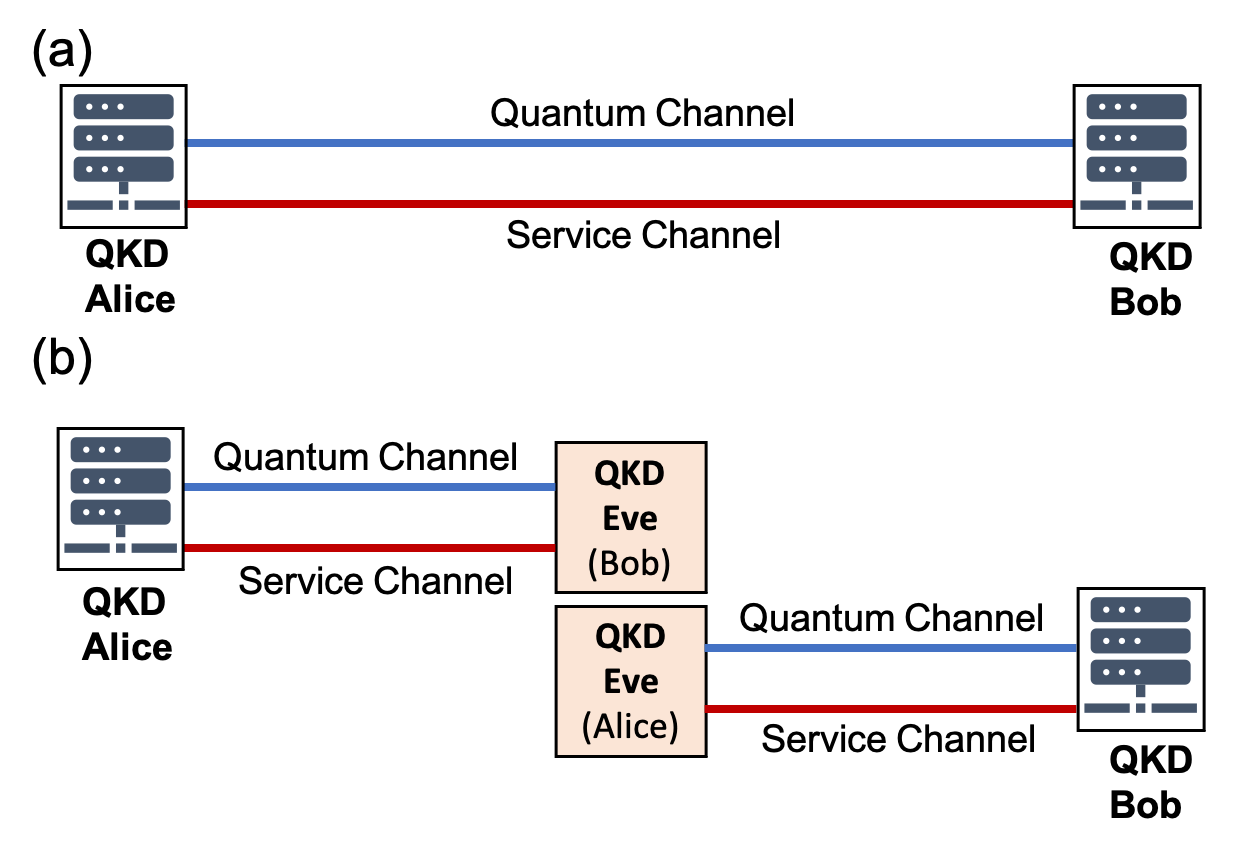}
\caption{Schematic representations of a 
point-to-point QKD link (a) and of the man-in-the-middle attack (b). }
\label{figure2}
\end{figure}

While processing their data, the two users have to verify the origin of the exchanged messages, otherwise the protocol is prone to a man-in-the-middle attack outlined in Figure \ref{figure2}. In this rather simple attack an adversary (Eve) cuts the quantum and the classical service channels 
that connect Alice and Bob. She connects her QKD devices to the loose ends of the channels, and pretends 
to be Bob to Alice and Alice to Bob. In this way, she can establish two secret keys, one with Alice and 
one with Bob, which allow her e.g., to decrypt and modify any message sent from Alice to Bob (or vice versa). Such an attack can be prevented only by authenticating the service channel, and given that QKD protocols promise information-theoretically  secure (ITS) distribution of random keys, 
authentication by means of ITS message authentication protocols is the natural choice for QKD \cite{PA95,Lutk99,Fung-etal10}. Standard ITS message authentication codes (MACs) \cite{book1,book2,book3,book4,Abidin2013,WC81,Kraw94}, however, require that  the two users share a secret truly random key, before they run their QKD system for the first time. 

For this reason, a typical commercially available pair (sender-receiver) of QKD devices comes with a pre-shared secret key, or the users who purchase the system, have to generate the key by some means, and transfer it to their devices. Among other tasks, this key is used for authentication in the first QKD session, until sufficient size of secret-key material has been produced. From that point on, message authentication for the subsequent QKD sessions is expected to consume part of the existing secret key. Clearly, a QKD protocol 
is meaningful only if  the key consumption is smaller than the  generation of 
fresh key in a session. For large distances and/or lossy channels, however, the  generation of fresh secret bits may become very costly, in terms of the amount of the pre-shared key 
required for their production. Another problem is that the need for a pre-shared secret key complicates considerably the design of large full-mesh QKD networks.  
These drawbacks, together with the strong dependence of QKD on  
ITS message authentication codes, make individuals, as well as 
public authorities and organizations to question the usefulness of QKD protocols,  and to keep a negative position with respect to further proliferation of the technology.  

The complications caused by the necessity for authentication 
have not gone unnoticed by the community, 
and there have been efforts in the past 
to decrease the length of the pre-shared key in existing QKD protocols,
as well as to fascilitate the distribution and the management of pre-shared keys. 
Peev {\em et al.} \cite{peev1} proposed an authentication scheme which relies on a two-step hash function evaluation, and involves a two-step non-ITS hashing, thereby 
resulting in low key consumption. Later, it was shown that the proposed scheme is not secure \cite{Abidin1,Pacher16}. 
More recently, Pan and co-workers \cite{Wang-etal21,Yang-etal21} studied experimentally the performance of QKD when the authentication relies on a post-quantum cryptographic (PQC) protocol \cite{Mosca-etal13}. Although this hybrid protocol has an efficient key 
management and new users can be easily added to the QKD network, it is not ITS, 
because the adopted PQC is computationally and not information-theoretically secure. 
In fact the key that is distributed by means of the hybrid protocol is secure provided that the underlying PQC has not been broken during the execution of the protocol, which can be guaranteed only in the framework of computational security.  Unless the adversary is present during the execution, it cannot later get hold of the distributed key, even if she has unlimited resources.

In the present work, our aim is to propose a scheme for the generation, the distribution and the management of pre-shared keys, which relies on physical unclonable functions (PUFs). 
By contrast to post-quantum public-key cryptosystems, the judicious integration  of PUFs in existing QKD systems, ensures ITS QKD under limited assumptions pertaining mainly to the performance of the PUF under consideration. Furthermore, our scheme allows for real-time 
authentication of the QKD devices that are connected to the network.  

PUFs were proposed by Pappu {\em et al} \cite{Pappu02}, as a means to generate random numerical keys from the laser speckle that is produced from a 
multiple-scattering optical medium. The optical medium plays the role of 
a token, which is technologically hard to clone and it can produce a large number of independent, almost truly random keys, each one associated with different parameters of the input laser light (e.g., wavelength, power, wavefront, angle of incidence, etc). 
Following the work of Pappu {\em et al}, PUFs have attracted considerable attention over the last two decades or so, and the related literature is rather rich \cite{PUF-review1,PUF-review2,PUF-review3,PUF-review4,PUF-review5,PUF-review6,PUF-review7,PUF-review8}. 
Nowadays, there are different types of PUFs, including optical (non-silicon) PUFs, time-delay based silicon PUFs,  intrinsic silicon PUFs, magnetic PUFs, etc. 
 Their main applications  are in entity authentication, and in the development of anti-counterfeiting methods, while there is also a number of companies, commercializing PUF products \cite{PUF-review2}.
For our purposes, it is sufficient to know the basic operation 
of a PUF, which is the generation of random keys, leaving aside details that go beyond the scope of the present work. The interested 
reader may refer to the related review articles 
\cite{PUF-review1,PUF-review2,PUF-review3,PUF-review4,PUF-review5,PUF-review6,PUF-review7,PUF-review8},  and the references therein.

\begin{figure}
\centering\includegraphics[width=9cm]{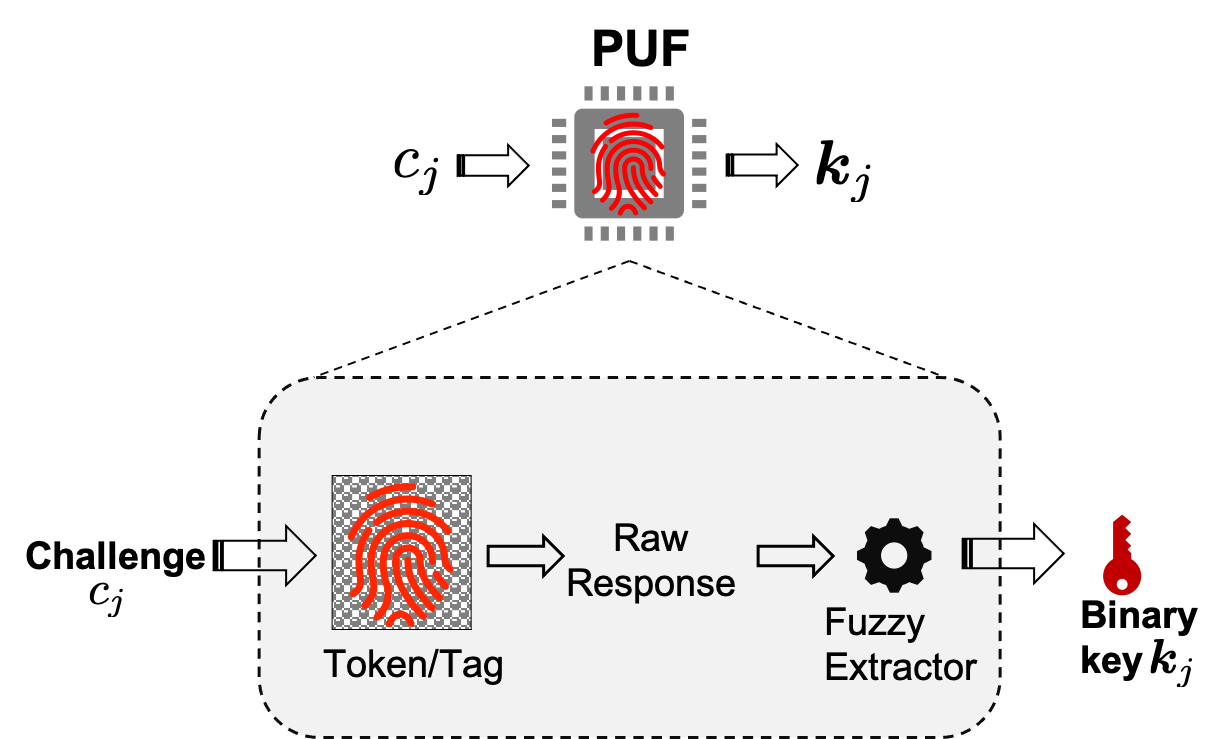} 
\caption{Schematic representation of a physical unclonable function (PUF). 
The token (sometimes also referred to as PUF tag), is a device with internal physical disorder. 
The internal disorder of the token is imprinted into its response to a physical challenge. The raw 
response is processed classically in order to yield a nearly perfect and robust random key.}
\label{figure3}
\end{figure}  

In a nutshell, as shown in Figure \ref{figure3}, a PUF is a mathematical function that refers to the behaviour of a physical object or device (to be referred to hereafter as the PUF token or tag). 
The PUF tag is characterized by internal random disorder, which plays the 
role of a fingerprint, and it is technologically hard to clone.  A PUF operates as a pseudo-random number generator (PRNG), in the sense that it produces a random numerical key  as a response to an input stimulus (to be referred to hereafter as challenge).  
The nature of the challenge 
depends on the PUF under consideration, while the response to a given 
challenge depends strongly on the internal disorder of the device. 
The raw noisy response is processed classically by means of a fuzzy extractor to yield a 
nearly perfect robust binary key. The fuzzy extractor typically involves two separate processes namely, 
reconciliation and hashing.  The former aims at a reconciled key which is not affected by environmental  variations and ageing of the token, whereas the latter compresses the reconciled key further, so that the final key is nearly uniformly distributed.   
A useful PUF is robust, easy to evaluate, difficult to replicate, unique (highly unlikely for two independent devices to return the same key for the same challenge), and unpredictable 
(it is very difficult or impossible to predict the response to a given challenge). Many of the PUFs that have been discussed in the literature satisfy these requirements. 
The main advantage of PUFs is that it removes the necessity of secret key storage, as the key is essentially physically stored in the disordered token, and can be recovered  
on demand when the appropriate challenge is given. The only requirement is that the user has  access to the  token. 

The strength of a PUF is generally determined by the number of supported challenge-response pairs (CRPs), or to make it more precise, by the way the number of potential CRPs scales with the increasing token size. In general,  a PUF for which the number of supported CRPs grows exponentially with the  size of the token is considered to be strong, while linear or polynomial growth typically refers to weak PUFs. Given that weak PUFs support a relatively small number of CRPs, 
an attacker can obtain the responses to all possible challenges, 
provided she has access to the PUF for sufficiently long period of time.  
Hence, she can emulate the particular PUF in any cryptographic protocol, without having an actual clone of the token. By contrast,  
strong PUFs support a much larger set of CRPs and an attacker has to have access to the PUF for a time period considerably longer than the one required for the readout  of a weak PUF. 
Ring-oscillator and SRAM PUFs are considered to be weak,  whereas optical and arbiter PUFs (a type of delay-based silicon PUF) are considered to be strong \cite{PUF-review1,PUF-review2,PUF-review3,PUF-review4,PUF-review5,PUF-review6,PUF-review7,PUF-review8}.

\section*{Results}
\label{section2}

We discuss now how PUFs can be integrated in existing QKD systems, facilitating the generation and the distribution of the necessary pre-shared key  between Alice and Bob.

\subsection*{Integration of PUFs in QKD systems}
\label{section2b}

The manufacturer of the QKD devices,  associates a PUF with each one of the QKD boxes 
(sender or receiver). For instance, a PUF tag is attached to each QKD box 
in a fashion similar to the barcodes or QR codes that accompany various products on the market. By contrast to these codes, however, as mentioned above the PUF tags are technologically hard to copy, and  they serve as  unique fingerprints for the boxes they are attached to. Alternatively, we can assume that the PUF token associated with a QKD device is given to the owner of the device 
e.g., in the form of smart card \cite{Nik21}.

\begin{figure}
\centering\includegraphics[width=9cm]{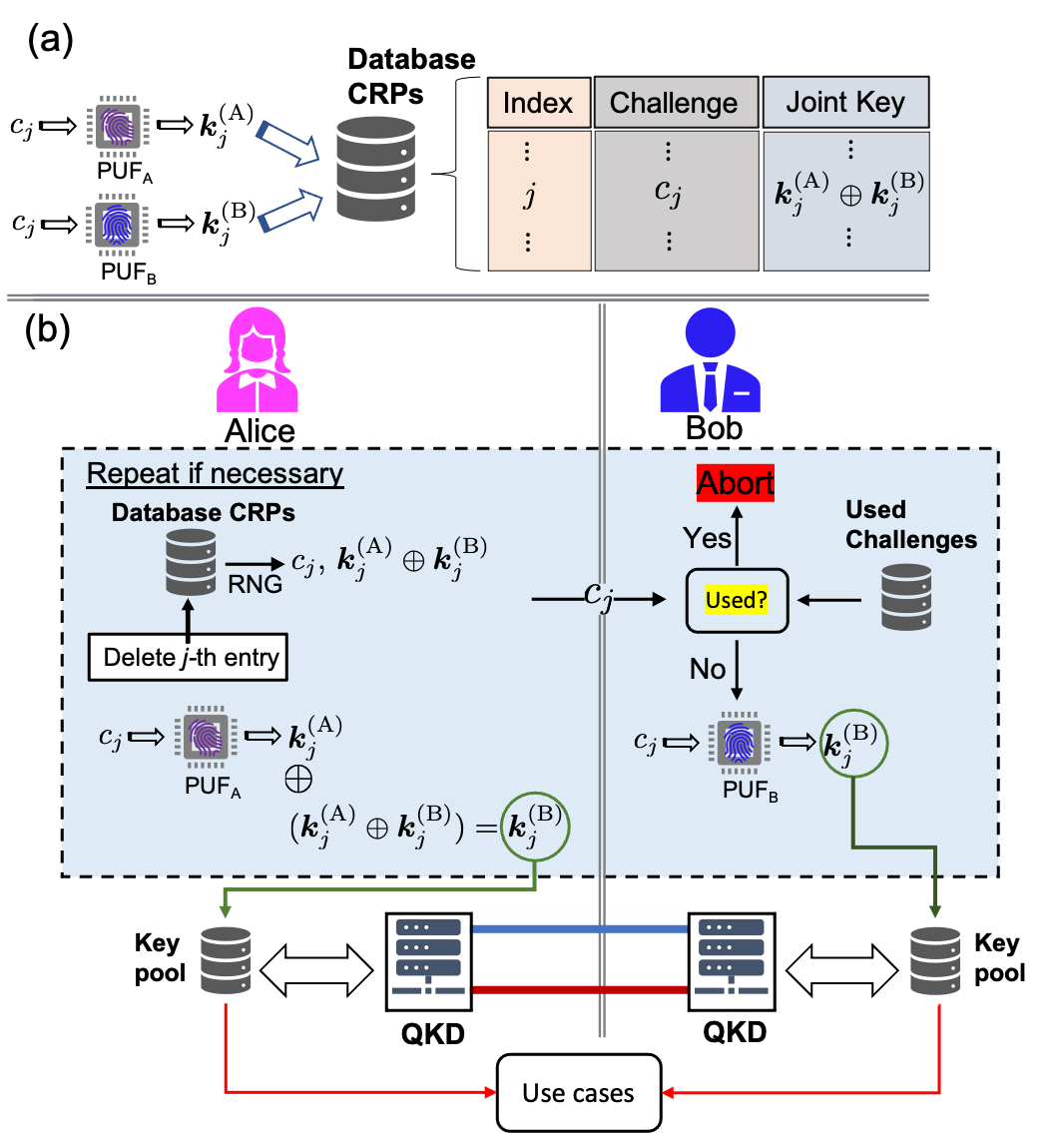} 
\caption{Integration of PUFs in a point-to-point QKD link. (a) Each pair of QKD boxes is associated with two PUFs namely, PUF$_{\rm A}$ and PUF$_{\rm B}$. A PUF generates a random key as a response to a challenge. The manufacturer creates a database of challenge-response pairs (CRPs), where only the joint keys are stored. (b) With the purchase of the QKD boxes, the users also have access to the corresponding PUFs. Moreover, one of them (say Alice), receives a copy of the database.  For the generation of a common 
random key, which will seed the first QKD session, Alice and Bob interrogate their PUFs independently with 
the same randomly chosen challenge. The corresponding entry is permanently removed from the database, while Bob also keeps track of the used challenges. This procedure can be performed again, e.g., if the first QKD session aborts, and a new QKD session is necessary. }
\label{figure4}
\end{figure}  
  
We will consider first a point-to-point communication scenario, with a 
pair of QKD boxes intended for Alice (sender) and Bob (receiver). 
A pair PUF tokens/tags is associated with them. 
Each token is interrogated by a number of challenges $\{c_1, c_2, \ldots\}$, and the corresponding keys $\{{\bm k}_1^{\rm (u)}, {\bm k}_2^{\rm (u)}, \ldots\}$ are recorded, where $u=\{{\rm A, B}\}$ 
is the label of the user. One of the honest users, say Alice, keeps a database of the challenge-response pairs (CRPs) in the following form \cite{Horstmayer13,Nik21} 
\bea
\begin{tabular}{cc|cc}
$c_1$& & & ${\bm k}_1^{\rm (A)} \oplus {\bm k}_1^{\rm (B)}$   \\
 $c_2$& & & ${\bm k}_2^{\rm (A)} \oplus {\bm k}_2^{\rm (B)} $  \\
 $\vdots$& & & $\vdots$ \\
 $c_j$& & & ${\bm k}_j^{\rm (A)} \oplus {\bm k}_j^{\rm (B)} $ \\
 $\vdots$& & &$\vdots$ \\
\end{tabular}
\eea
A schematic representation of the process is shown in Figure \ref{figure4}(a).  As will be discussed 
in detail below, the particular form of storage protects the individual keys against an adversary who obtains undetected access to the  database.  

The question arises here is who  creates and who  has access to this database. 
There are two options. One possibility is that the manufacturer creates the database 
and gives it to either of the two honest users during the delivery of the QKD boxes. 
Alternatively, one of the honest users (say Alice), has access to both boxes (and thus to both PUFs) simultaneously, and she creates the database before she gives one of the boxes to Bob. In both of these 
scenarios, only trusted parties are involved, in the sense that by definition Alice, Bob and the manufacturer 
of the QKD devices are honest. In Figure \ref{figure4}, we assume that Alice possesses the database and hence there is no need for including an additional  trusted third party. As will be discussed in the next subsection, however, the  inclusion of a third party is inevitable in the case of larger QKD networks. 

As discussed in the introduction, before Alice and Bob operate their QKD boxes for the first time, they have to generate a common secret key, which will be used for post-processing purposes, until sufficiently large amount of secret bits has been generated from the QKD sessions. To this end, Alice chooses at random
one of the available challenges in the database, say the $j$th one, 
and sends it to Bob  [see Figure \ref{figure4}(b)]. The two users run their PUFs independently in order to recover their individual keys 
${\bm k}_j^{\rm (A)}$ and ${\bm k}_j^{\rm (B)}$. Alice adds her key to the joint key 
${\bm k}_j^{\rm (A)} \oplus {\bm k}_j^{\rm (B)} $  in order to recover 
Bob's key which will seed the first QKD session. The corresponding entry is permanently deleted from the database, while Bob also keeps a blacklist of the used challenges, so that they are 
not used again for the particular PUFs. 

Under normal conditions, the key material in the corresponding pools of Alice and Bob, will grow through the running QKD sessions, and there will be no need 
for the users to execute the above procedure again. However, if for any reason, at any time, QKD alone cannot support the needs for keys, and additional key material is needed, the procedure can be repeated again.     

\begin{figure}
\centering\includegraphics[width=8.6cm]{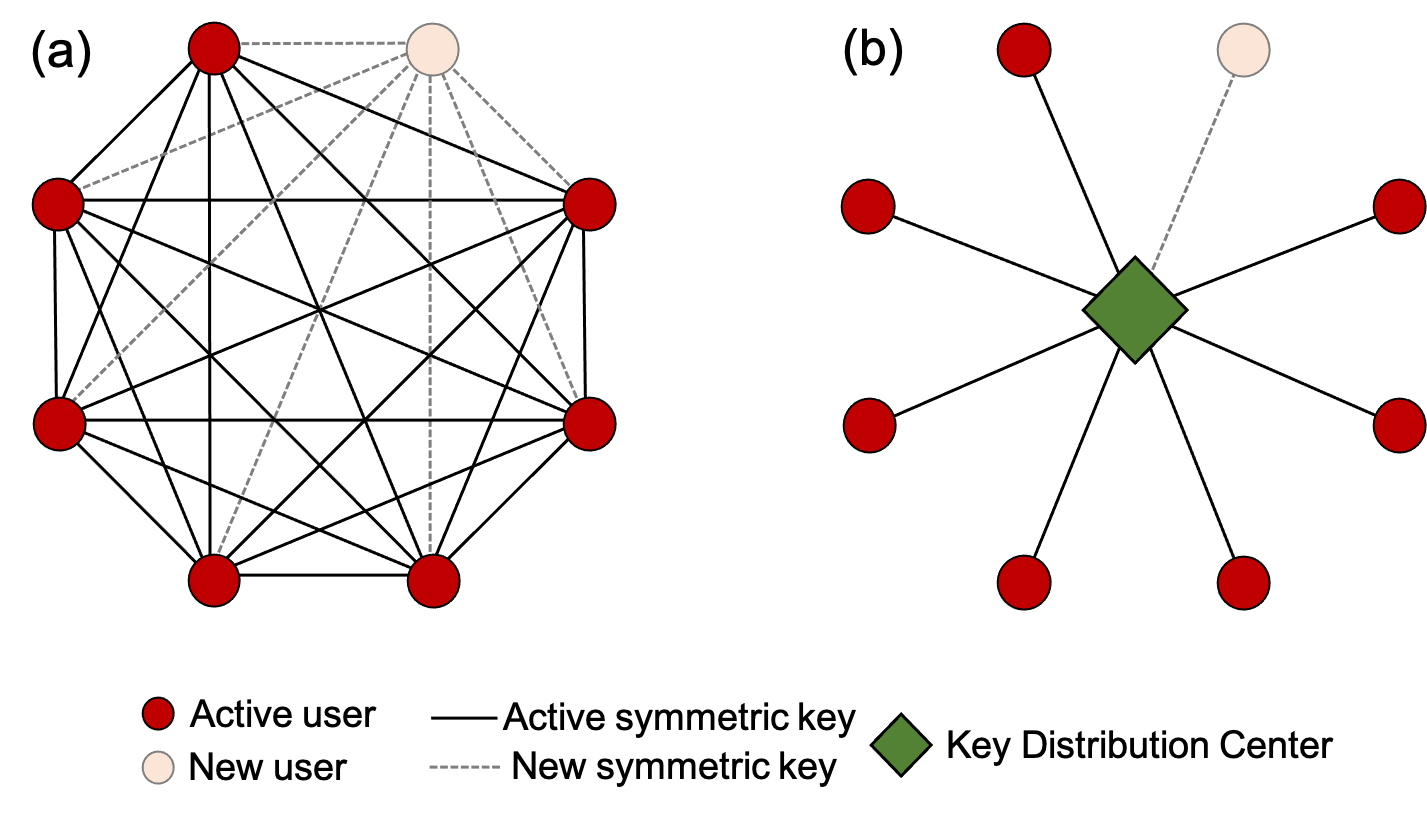}
\caption{Full-mesh QKD network involving $n$ users.   
(a) In the absence of a key distribution center (KDC), 
the total number of pre-shared keys is $n(n-1)/2$, 
while each new user has to share $n$ keys with each one of the other existing  users. 
(b) In the presence of a KDC, each user shares a key with the KDC only.}
\label{figure5}
\end{figure}


\subsection*{Large QKD networks}
\label{section2c}

The previous subsection focuses on a typical point-to-point QKD scenario, where only two honest parties are involved, and a single pre-shared key is required for the seeding of the first QKD session. 
The design of large full mesh QKD networks with $n$ users, however, requires  
$n(n-1)/2$ pre-shared keys [see Figure \ref{figure5}(a)].  
For large $n$ this is a big amount of key material 
that has to be distributed in a secure manner among the users, in order for the 
first pairwise QKD sessions to be implemented. 
Moreover, it is really hard for a new user to be added in the network. 
Indeed,  as soon as a new user (let us say Charlie) receives his QKD device, 
this should come with a list of keys pertaining to each one of the other users with whom Charlie may 
want to establish a QKD link. Each one of these users has to be informed about Charlie joining the network,  and he/she should receive a copy of the corresponding symmetric key to be used for the implementation of the 
first QKD session with him. Overall, the addition 
of a new user in the network with $n$ users, should be accompanied by the generation and the distribution of $n$ additional symmetric secret keys; a rather tedious task.

One solution to this problem is the use of a trusted center, which shares a secret key with each one of the $n$ users, thereby reducing significantly the 
number of pairwise shared  secret keys [see Figure \ref{figure5}(b)]. In this case, it is relatively easy for Charlie to join 
the existing network, as he has to obtain together with his 
QKD device, only one secret key which is known to the key-distribution center (KDC).  The establishment of a QKD link with any other user 
can be achieved through the KDC, which can help the users to agree on a common secret key required 
for the post-processing in the first QKD session. The second solution to the aforementioned problems 
relies on the use of classical or quantum asymmetric cryptosystems, where each user holds a pair of keys (a private key and a public key). Standard, widely used  asymmetric cryptosystems rely on the hardness of certain 
mathematical problems, and thus they offer computational security only \cite{Mosca-etal13,Wang-etal21,book1,book2,book3,book4,PQ1,PQ2}. 
On the other hand, a small number of ITS quantum public-key cryptosystems have been proposed in the literature, but their implementation is beyond reach of current technology \cite{NikPRA08,Kaw05,Kab00}. 

As long as we are interested in preserving the ITS character of QKD protocols, the only currently available route towards the realization of large  QKD networks involves one way or another the use of a trusted authority, such as a KDC,  
which is trusted by all the parties participating in the network. 
This solution is by no means optimal, because the functionality and  
the security of the entire QKD network depends strongly on the KDC, which 
automatically becomes the main target of potential adversaries. 
Given the current status of quantum technologies, however, 
the presence of a KDC in large networks facilitates and guarantees 
the generation and the distribution of  pre-shared keys between users, which are essential for the establishment of the first secure pairwise QKD sessions.  
In the literature of QKD, the inclusion of a KDC is also considered essential for one more reason namely, it can operate as a relay thereby allowing two users that are  separated by a distance well beyond the reach of current QKD systems, to establish a secret key \cite{qkd_review6}. 

\begin{figure}
\centering\includegraphics[width=9cm]{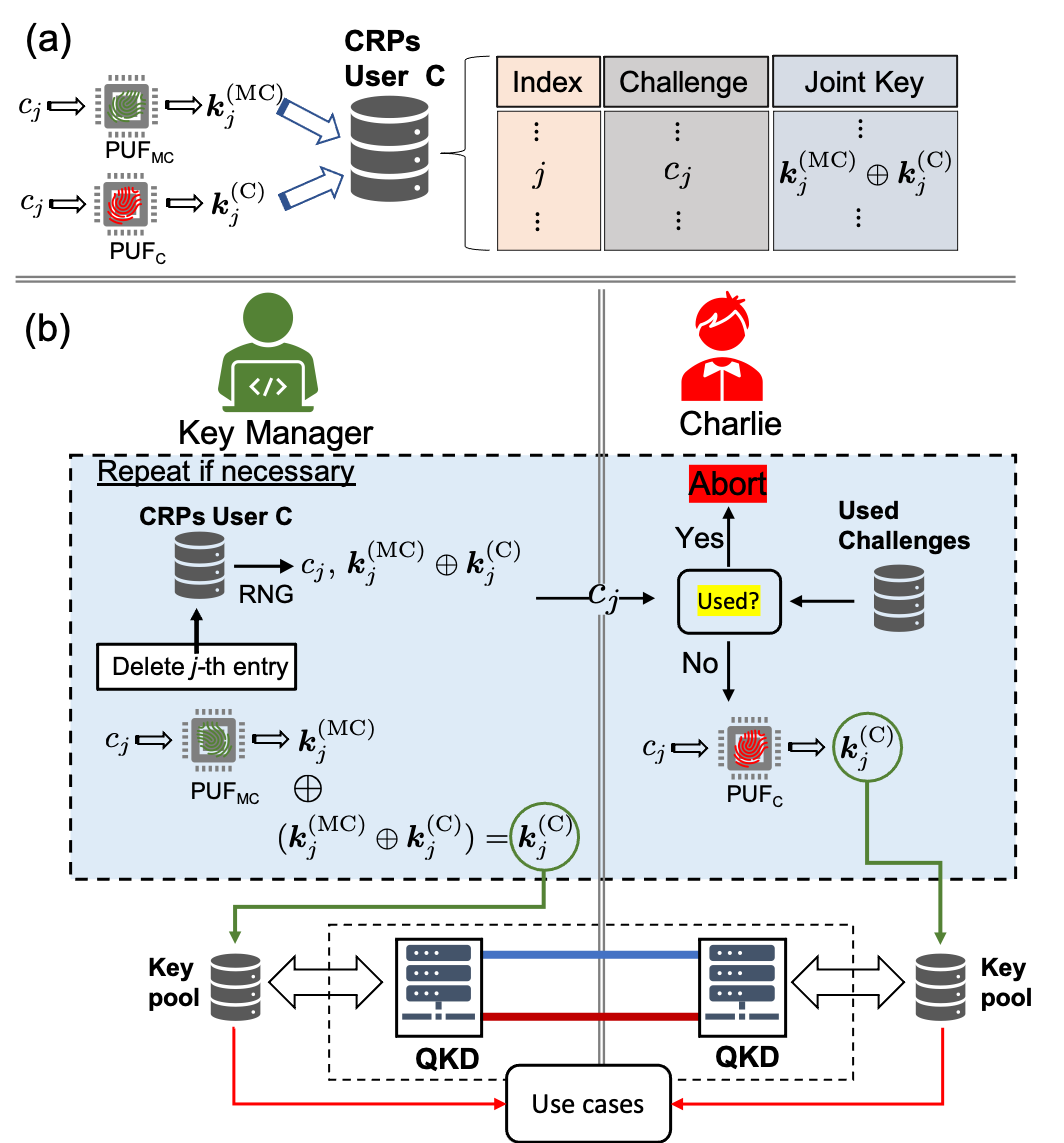} 
\caption{Integration of PUFs in a large QKD network. (a) Each QKD device (sender or receiver) is associated with a PUF. The PUF generates a random key as a response to a challenge. The KDC has its own PUFs, and it is controlled by the  manufacturer, who creates a database of challenge-response pairs for each QKD device. 
The KDC has access to the databases of CRPs for all of the QKD devices that have been or will be connected to it, while a different PUF is used for the encryption of the entries in each database. 
(b) Each time fresh key material is needed for a user, the user can generate a common 
random key with the KDC,  by running the protocol in the shaded box. Note that the presence of the QKD link between the KDC and the user is not necessary, 
as keys can be generated from the PUFs. In practice, the addition of such a QKD link will limit the distance  between the user and the KDC, but it can make the network self sustainable in terms of key generation and consumption.}
\label{figure6}
\end{figure}

The ideas discussed in  the previous subsection can be generalized in 
the context of large QKD networks,  if we assume that the KDC 
holds all the databases of CRPs, as shown in Figure \ref{figure6}. 
The manufacturer assigns a PUF tag to each QKD device (sender or receiver), and a separate database 
of CRPs is created along the lines discussed in the previous subsection. 
For the sake of simplicity, let us assume that the manufacturer of the QKD devices also controls 
the KDC, while the  manager of the KDC has one PUF for each database i.e., for each user. 
So, the KDC manages as many databases and PUFs, as the user devices in the QKD network. 
For the sake of concreteness, 
let us consider the database for the QKD device C, intended for Charlie who will join the QKD 
network. Let PUF$_{\rm MC}$ denote the PUF used by the KDC, for the encryption of the entries 
in Charlie's database of the CRPs. The $j$th entry in the database involves the challenge $c_j$ and the 
joint key ${\bm k}_j^{\rm (MC)}\oplus {\bm k}_j^{\rm (C)}$, where ${\bm k}_j^{\rm (MC)}$ 
and ${\bm k}_j^{\rm (C)}$ denote the keys associated with the responses of PUF$_{\rm MC}$ and 
PUF$_{\rm C}$ to challenge $c_j$, respectively.  
As soon as Charlie is ready to join the network, his first task is to establish a common secret key 
with the KDC,  following the steps discussed in the previous subsection. The procedure is 
recapitulated in Figure \ref{figure6}(b), with the necessary changes in the labels. 

\begin{figure}
\centering\includegraphics[width=15cm]{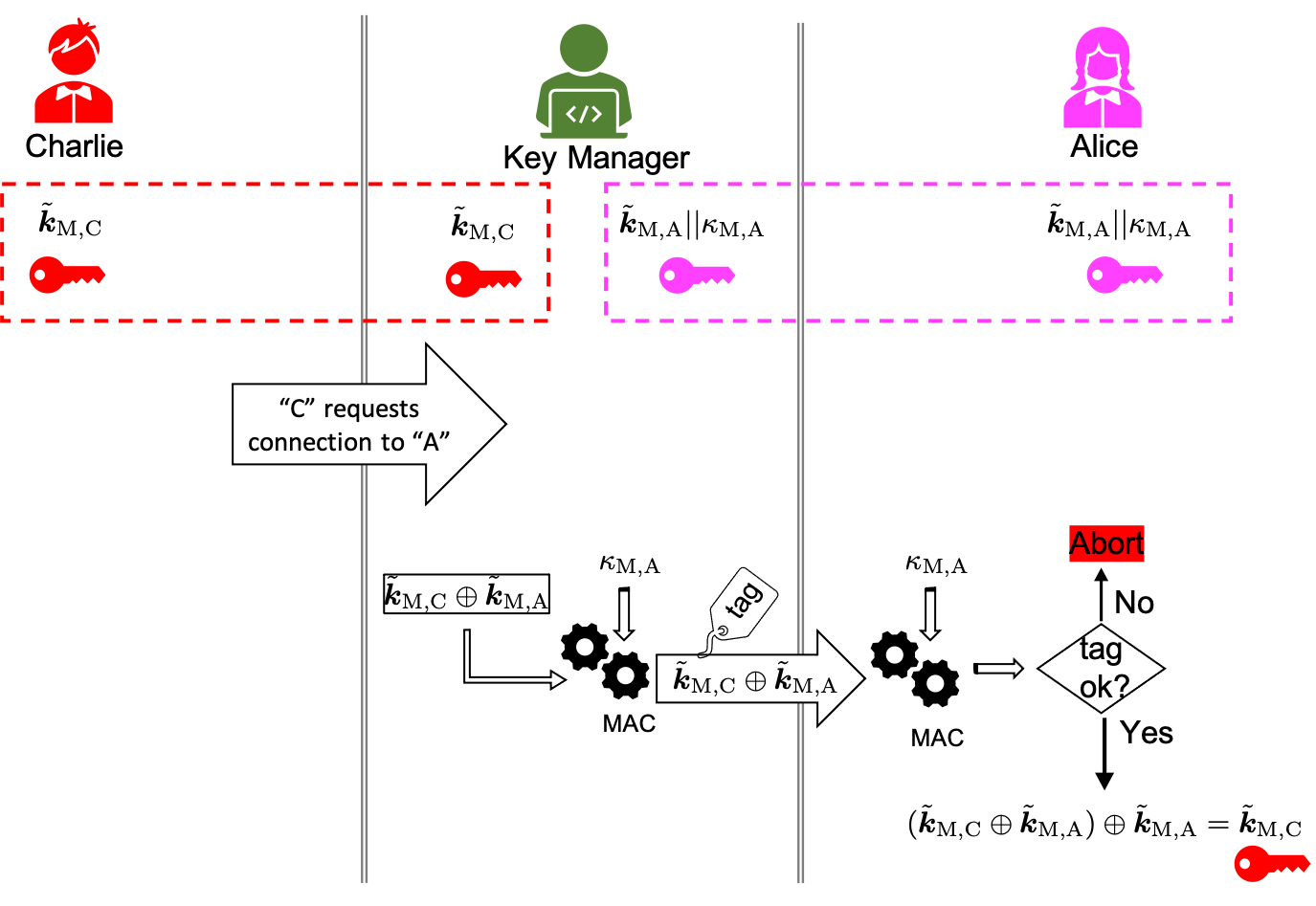}
\caption{Schematic representation of the procedure through which Alice and Charlie 
can establish a common secret key, with a third party acting as a relay. }
\label{figure7}
\end{figure}

Having established a common secret key with the KDC, Charlie can establish a common secret 
key with any other existing user in the network, which will allow them to run the  post-processing in the first QKD sessions, until sufficiently large amount of fresh secret random bits has been generated through the QKD. 
The procedure is outlined in Figure \ref{figure7} and it is as follows. 
The key manager of the KDC shares a common secret key 
$\tilde{\bm k}_{\rm M,C} $ with Charlie, and another key with Alice, a user who is already in the 
network, and with whom Charlie wants to communicate. For the sake of clarity, we write the latter key as a concatenation of two smaller keys i.e., 
$\tilde{\bm k}_{\rm M,A} || {\bm \kappa}_{\rm M,A}$. 
Charlie contacts the KDC and requests connection with Alice, who is already in the network. The key manager calculates 
$\tilde{\bm k}_{\rm M,C} \oplus \tilde{\bm k}_{\rm M,A}$, and sends it to Alice, over a classical channel. The message is authenticated with the key ${\bm \kappa}_{\rm M,A}$. 
Upon reciept of the message, Alice confirms its origin, and she  adds to the received message her key to obtain 
$\tilde{\bm k}_{\rm M,A} \oplus (\tilde{\bm k}_{\rm M,C} \oplus \tilde{\bm k}_{\rm M,A}) =\tilde{\bm k}_{\rm M,C} $.  Hence, Alice and Crarlie share a common secret key, which can be expanded through QKD. 

It is worth emphasizing here, that the realization of the ideas we have just presented does not require a QKD link between each user and the KDC. 
As shown in Figure \ref{figure6}, the manager and the user can agree on a common  secret 
key by means of classical communication and local operations on their PUFs. The only requirement is for the manager to 
communicate a randomly chosen challenge to Charlie, over a public channel. This allows a lot of flexibility with respect to the physical topology of the network, in the sense that there are no limitations on the spatial separation between a user and the KDC, which are associated 
with the transmission of quantum signals. 
In the absence of QKD links between the KDC and the users, all the key material is provided only by the PUFs, and thus it is limited by the number of challenge-response pairs that can be supported by the PUFs under consideration. In this context, strong PUFs are preferable. 
However, although a QKD link between Charlie and the KDC imposes limitations on the distance, at the same time it allows for the expansion of the initial key that has been obtained through PUFs. In this case the network may become self-sustainable, with respect to the generation and the consumption of key material, and its operation is not expected to  depend strongly on the type of used PUFs.

\subsection*{QKD device authentication}
\label{section2d}

Entity authentication (also known as identification), is a very important cryptographic task, which allows the identity of a user or device to be confirmed \cite{book1,book2,book3,book4}. By means of entity authentication one can control which users or devices have access 
to certain physical or virtual resources. Entity authentication should not be confused with message authentication. They are  different cryptographic tasks with distinct goals, and one cannot replace the other. A fundamental difference between the two, is that an identification scheme provides  real-time evidence about the identity of a user or a device, 
whereas MACs  (and signature schemes) allow for data origin authentication, which can be performed any time after the relevant message has been tagged or signed.

The integration of PUFs in QKD networks allows the KDC to confirm at any time 
the identity of the QKD devices that are connected to the network, thereby preventing 
the connection of counterfeit or unauthorized devices, through which an attacker may 
try to connect to the KDC or to other users in the network, 
in order to access databases, or to sabotage the operation of the network.  
Moreover, an honest user can confirm the authenticity of a purchased QKD device. 
A PUF tag that is attached to a QKD device, serves as a unique fingerprint, and protects 
against counterfeiting.

\begin{figure}
\centering\includegraphics[width=0.5\textwidth]{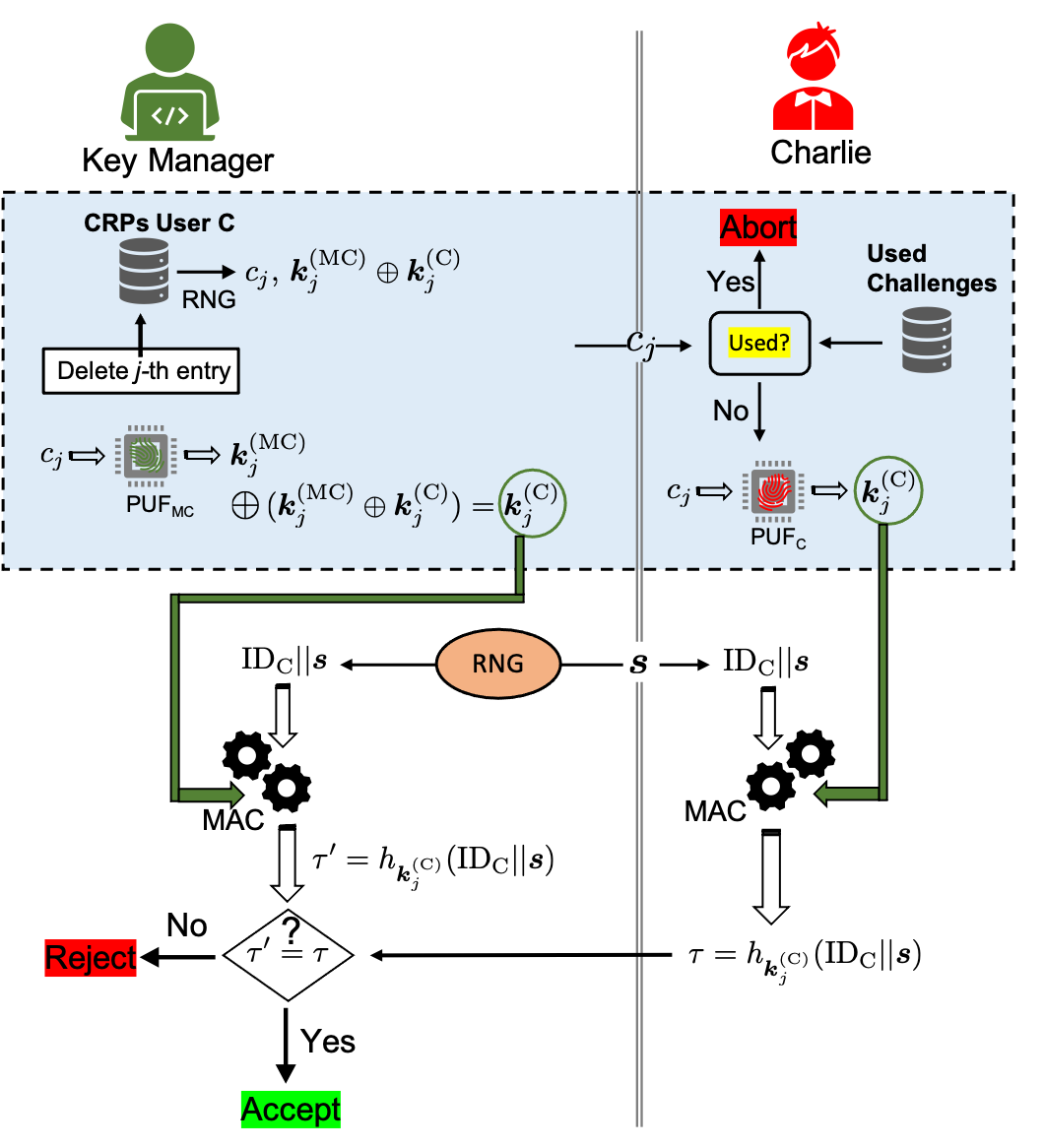}
\caption{Schematic representation of an entity authentication session. 
In order for the KDC to confirm the identity of the newly connected QKD device of Charlie, 
they run the protocol 
in the shaded region, to obtain a common secret key through their PUFs. 
The key manager also chooses a random binary string ${\bm s}$ and he sends it to the user. 
The user runs a publicly known MAC to obtain a tag $\tau$ for the 
concatenated message involving his unique identity (in binary format) ID$_{\rm C}$,  
and the received random binary string. The tag is sent to the KDC where it is compared 
to the tag produced locally by the key manager, and the identity of the QKD device is accepted  
only if the two tags agree. If a secret key is already shared between the user and the KDC, 
then the steps in the shaded box can be omitted.}
\label{figure8}
\end{figure}

Let us consider the scenario where a new user, Charlie, joins the network, and that the KDC 
wants to authenticate his QKD device. The procedure is outlined in Figure \ref{figure8}.
As in the previous subsections, we will assume that Charlie's QKD device is associated with 
PUF$_{\rm C}$. The related database of CRPs is generated by the manufacturer and it 
is available to the KDC. Moreover, we will assume that each user (or QKD device) is uniquely identified by a binary string (a serial number). 
Given that Charlie is a new user, who does not share a secret 
key with the KDC, the key manager chooses at random one of the available challenges from the database of CRPs associated with  PUF$_{\rm C}$.  
Moreover, the manager generates  a random bit string ${\bm s}$, and he sends 
it to Charlie together with the randomly chosen challenge $c_j$.  Charlie obtains the response 
${\bm k}_j^{\rm C}$ 
of PUF$_{\rm C}$ to challenge $c_j$, whereas the  manager recovers the same response after he adds to the joint key, the  response of PUF$_{\rm MC}$ to the same challenge. 
Charlie passes the concatenation of his identification string 
${\rm ID}_{\rm C}$ with ${\bm s}$ through a MAC with key ${\bm k}_j^{\rm C}$, and 
the resulting tag  
$\tau = h_{{\bm k}_j^{\rm C}}({\rm ID}_{\rm C}||{\bm s})$, 
is sent to the KDC. 
The key manager computes the tag for the same message and he accepts the QKD device as 
authentic, only if this agrees with the tag received from Charlie. 

In the protocol outlined in Figure \ref{figure8}, we have assumed that the user does not share a 
common secret key with the KDC and it has to be generated locally through their PUFs. 
If the user shares a secret key with the KDC, then the procedure in the shaded region of 
Figure \ref{figure8} can be omitted, and the entity authentication can rely on the existing key material. 
Moreover, the protocol can be easily extended to mutual entity authentication, by adding a second round of authentication, where the user chooses at random another binary string,  
which is sent to the KDC \cite{book1,book3}.


\section*{Security considerations} 
\label{section3}

The Wegman-Carter authentication (WCA) scheme that is usually employed in standard QKD systems, is ITS only if the pre-shared secret key is truly random, or at least very close to perfect.  
In particular, it has been shown that a WCA scheme 
which is executed with an $\epsilon^\prime-$perfect key, is indistinguishable from the ideal scenario, with probability $(1-\varepsilon-\epsilon^\prime)$, 
where $\varepsilon$ refers to the security of the ideal scheme \cite{Abidin2013,AbiLar2014}.   The protocols outlined in Figures \ref{figure4}, \ref{figure6} and \ref{figure7}, can provide two honest users with 
the secret key that is necessary for an ITS post-processing in the first QKD session, if the following conditions are satisfied:
\begin{enumerate} 
\item[(C1)] The numerical keys produced by the PUFs under consideration are close to truly random.
\item[(C2)] The legitimate users never make their PUF tokens/tags available to other parties. 
\item[(C3)] Each entry in the database of challenge-response pairs is used only once. 
\item[(C4)] The MAC that is used for the distribution of the key in Figure \ref{figure7} is ITS.  
\end{enumerate} 

For various PUFs that have been discussed in the literature, the generated keys have been shown to pass successfully widely accepted tests 
of random sequence certification, such as the ones provided by the NIST suit \cite{Pappu02,PUF-review6,Mes18,Horstmayer13}. 
For all practical purposes such a key can be considered to be close to truly random, 
and there are no  correlations between different elements or parts of the key. 
Moreover, keys  that have been generated from different PUFs 
or from the same PUF but for different challenges, can be considered as 
uniformly distributed independent random strings. 
As a result, the entries in a database of CRPs are ITS, because 
the extraction of the individual keys ${\bm k}_j^{\rm (u)}$ and ${\bm k}_j^{\rm (v)}$ from the joint key ${\bm k}_j^{\rm (u)}\oplus {\bm k}_j^{\rm (v)}$ is impossible 
by virtue of the one-time-pad (OTP) encryption, 
unless the attacker has also access to at least  one of the PUFs. 
Furthermore, an attacker cannot launch a machine learning attack \cite{PUF-review2}, in order to create a model of either of the two PUFs. This is because machine learning requires direct access to a set of CRPs $\{(c_j , {\bm k}_j^{\rm (u)}) : j=1, 2, \ldots \}$, 
whereas in our case the individual keys are not accessible and they are stored encrypted in a secure manner.

For the sake of simplicity, in Figures \ref{figure4} and \ref{figure6} we have assumed that 
one challenge suffices for the seeding of the first QKD session. This may not be always the case.  
The length of the key that can be generated from a PUF for a single challenge ranges from a few hundreds to a few thousands of bits. The precise length depends on the type of the PUF under consideration, as well as on the details of the fuzzy extractor. If a single challenge cannot provide a key of the appropriate length for seeding 
the first QKD session, then one can concatenate two or more keys, pertaining to different challenges.
Indeed, as long as different challenges yield independent nearly perfect keys,  the  resulting longer key is also close to truly random.  

Let us turn now to the entity authentication protocol of Figure \ref{figure8}. 
One can readily show that it is information theoretically secure if an $l$-time $\varepsilon$-secure MAC is used\cite{book1,book3,NikFis00}, with $l\ll 2^{\mid{\bm s}\mid}$. 
In particular, the probability that the KDC uses the same random string ${\bm s}$ 
in the $i$th session is $(i-1)/2^{\mid{\bm s}\mid}$. Hence, the probability for a replay in any 
of the $l$ sessions is $l(l-1)/2^{\mid{\bm s}\mid+1}$. The same result can be obtained through the 
Birthday bound, for $l\ll 2^{\mid{\bm s}\mid}$. On the other hand, there is 
probability at most $\varepsilon$ for an adversary to construct the correct tag. Hence, the overall 
probability for an adversary to deceive the KDC is at most 
$l(l-1)/2^{\mid{\bm s}\mid+1}+ \varepsilon$. Of course, one should not forget here that unconditionally secure $l$-time $\varepsilon$-secure MACs, require a uniform truly random key. 
Hence, the previous discussion on the quality of the keys that are produced from PUFs, and the security of the databases also affects the security of the entity authentication protocol of Figure  \ref{figure8}.

Another issue that deserves our attention is the communication of the randomly chosen challenge 
$c_j$ in the procedures outlined in Figures \ref{figure4}(b), \ref{figure6}(b) and \ref{figure8}. 
The challenge is sent in plain text over a public, possibly unauthenticated, 
channel. An adversary who learns the challenge, and he does not have access 
to the PUF of the sender or the receiver, cannot use it to her advantage, because as discussed 
above, the PUFs operate as PRNGs, and their output to a given challenge cannot be predicted with probability much better than random guessing. 
Even if the adversary has access to the database of CRPs, she cannot deduce the individual keys 
from the joint key, by virtue of the OTP encryption. Finally, given that the entry pertaining to the communicated challenge is deleted from the database of CRPs and is added to a blacklist, an adversary cannot replay it, so that to fool any of the two users.
What the adversary can do, is to launch some sort of DoS attack, by changing the challenge. In this case the two users will essentially not share the same key, as required for successful realization of a QKD session, and it is almost certain that the different keys will make the QKD protocol to abort, and the users will have to start all over again. 
A similar situation we have in Figure \ref{figure7}, where  Charlie's request for communication with Alice is also sent in plain text, and its origin is not verified by the receiver. In principle, an attacker can send multiple such requests to the KDC, thereby enforcing the manager to consume keys 
without reason, and depleting the relevant databases and key pools. 
In all of these scenarios, the attacks are possible because the classical communication is not authenticated. However, none of these attacks threatens the security of the QKD network, 
but rather they aim at the resources of the QKD network. For this reason, there is no need for ITS message authentication, and they can be prevented by means of computationally secure MACs 
\cite{book1,book2,book3,book4,BKR00} or public-key cryptosystems \cite{book1,book4,PQ1,PQ2}.  

Finally, throughout this work we have assumed that the manager 
of the KDC uses a different PUF for each user participating in the network, namely 
PUF$_{\rm MA}$, PUF$_{\rm MB}$, PUF$_{\rm MC}$, etc. 
In this way, we have added an additional level of security, in the sense that if, for any reason, 
a database of CRPs is compromised, the security of the other databases is not affected, as long as different PUFs produce independent random keys. In particular,  the $j$th entry in the databases of CRPs for two different users,  say A and B, pertains to the joint keys $K_1:={\bm k}_j^{{\rm (MA)}}\oplus {\bm k}_j^{\rm (A)}$ and $K_2:={\bm k}_j^{{\rm (MB)}}\oplus {\bm k}_j^{\rm (B)}$, where all of the individual keys refer to the same challenge but to different independent PUFs, namely 
PUF$_{\rm MA}$, PUF$_{\rm A}$, PUF$_{\rm MB}$ and PUF$_{\rm B}$.  
By virtue of the OTP encryption, even if the user A is malicious, she 
cannot deduce either ${\bm k}_j^{{\rm (MB)}}$ or ${\bm k}_j^{\rm (B)}$ from 
$K_1$ and PUF$_{\rm A}$, provided that conditions (C1) and (C2) are satisfied for the PUFs under consideration.


\section*{Discussion} 
\label{section4}

We have considered the distribution of a common secret key between two honest  users of a QKD network, which is necessary for authentication and encryption purposes during the post-processing stage in the first pairwise QKD session. Such a key (usually referred to as pre-shared key in the literature of QKD), is necessary for any QKD system, because it prevents a man-in-the-middle attack. 

We have proposed the generation and the distribution of pre-shared keys by means of PUFs.  
Each QKD device (sender or receiver), is associated with a disordered token/tag, which plays the 
role of a unique fingerprint, and allows the users to generate nearly perfect random keys, as well 
as to authenticate their devices. Two users can generate a common secret random key, by 
interrogating locally their tokens with the same randomly chosen challenge. The procedure does not 
require quantum communication, it offers information-theoretic security, and it can be extended 
to large QKD networks with the use of a trusted KDC. 

As discussed above, the inclusion of a trusted KDC is essential in large QKD 
networks, in order to preserve the ITS character of QKD systems. If one is willing to accept a computationally secure  system, PUFs can be combined with standard public-key cryptosystems in order to generate and distribute keys between users. In particular, the key that is generated from a PUF device may serve 
as a private key, or it can be used as a random seed for the generation of a private key. 
Subsequently, building upon the private key,  the corresponding public key is generated, which becomes publicly available through a trusted public-key server. The advantage of such an asymmetric cryptosystem is that it opens up the way for the implementation of various cryptographic tasks beyond the generation and the distribution of a key, including digital signatures. At the same time, the pair of the private-public keys is 
bound to the PUF device. 

Throughout this work we have adopted a rather general theoretical framework, which is not 
restricted to a particular type of PUF. The present results pave the way for the integration 
of PUFs in currently used QKD systems, but there are many practical issues that deserve 
further investigation, in order to identify what type of PUF serves better the needs of QKD networks. 
To this end one has to take into account various facts including the required key lengths, 
the size of the network, the robustness of the PUF in the presence of environmental fluctuations, etc. 
Optical PUFs may be a very promising candidate, because they are considered 
among the strong PUFs, and typically they can support long keys (thousand of bits) 
\cite{Pappu02,Mes18,Horstmayer13}. Moreover, they are  compatible with the QKD 
infrastructure and, in principle, they are amenable to remote quantum readout \cite{Skoric12}. Remote quantum readout of optical PUFs is very attractive if one is interested 
in saving classical resources, at the cost of introducing limitations in the spatial separation 
between the users and the KDC. However, all the known quantum readout schemes \cite{Goorden14, NikDiaSciRep17, Nik18} are currently 
limited to very short distances (of the order of 1 km), and there is need for 
their extension to distances comparable to the ones that can be covered by standard QKD systems. 
We believe that experimental research may shed light on the integration of PUFs in operational 
QKD networks, and provide answers to many of these questions.


\section*{Data availability}

All data generated or analysed during this study are included in this published article (and its supplementary information files).

\section*{Acknowledgements}

This work has been funded by the Deutsche Forschungsgemeinschaft (DFG, German Research Foundation) – SFB 1119 – 236615297.
The authors are grateful to Prof. G. Alber, for useful discussions. 

\section*{Author contributions statement}

GMN  conceived the main idea and developed the theory together with the security analysis. 
MF contributed to the security analysis as well as to practical aspects pertaining to the implementation of the protocols.  Both authors reviewed the manuscript. 

\section*{Additional information}

The authors declare no competing financial interests.


\end{document}